\documentclass[floatfix,nofootinbib,superscriptaddress,twocolumn,pre,showpacs,showkeys,preprintnumbers]{revtex4-1}
\usepackage{amsmath,amssymb,amsthm,amsfonts}
\usepackage{epsfig,graphics,color,calc,graphicx}
\usepackage{subfigure}
\usepackage{comment}
\usepackage{pgf,tikz}
\usepackage{mathrsfs}
\usepackage{color}
\usetikzlibrary{arrows}
\usepackage{mathrsfs}
\usepackage{hyperref}
\hypersetup{colorlinks=true,linkcolor=blue}
\usepackage{mathrsfs}

\newlength{\pecettawidth}
\begin{document}
\title{Residence time of symmetric random walkers in a strip with large 
reflective obstacles}

\author{Alessandro Ciallella}
\email{alessandro.ciallella@uniroma1.it}
\affiliation{Dipartimento di Scienze di Base e Applicate per l'Ingegneria, 
             Sapienza Universit\`a di Roma, 
             via A.\ Scarpa 16, I--00161, Roma, Italy.}

\author{Emilio N.M.\ Cirillo}
\email{emilio.cirillo@uniroma1.it}
\affiliation{Dipartimento di Scienze di Base e Applicate per l'Ingegneria, 
             Sapienza Universit\`a di Roma, 
             via A.\ Scarpa 16, I--00161, Roma, Italy.}

\author{Julien Sohier}
\email{julien.sohier@u-pec.fr}
\affiliation{Laboratoire d'Analyse et de Math\'ematiques Appliqu\'ees (LAMA),\\ 
UMR 8050, Universit\'e Paris Est Cr\'eteil,\\
61, avenue du G\'en\'eral de Gaulle, 94010 Cr\'eteil Cedex, Paris, France.}


\begin{abstract}
We study the effect of a large obstacle on the so called
residence time, i.e., the time that a particle performing a symmetric 
random walk in a rectangular (2D) domain needs to cross the strip. 
We observe a complex behavior, that is we find out that the residence 
time does not depend  monotonically on the geometric properties of 
the obstacle, such as its width, length, and position. 
In some cases, due to the presence of the obstacle,
the mean residence time is shorter
with respect to the one measured for the obstacle--free strip. 
We explain the residence time behavior by 
developing a 1D analog of the 2D model where the role of the 
obstacle is played by two defect sites having a smaller probability to be 
crossed with respect to all the other regular sites. 
The 1D and 2D models behave similarly, but
in the 1D case we are able to compute exactly the residence time 
finding a perfect match with the Monte Carlo simulations. 
\end{abstract}




\maketitle

\section{Introduction} \label{ENMC:sec:introduction}
\par\noindent
When a particle flow crosses a region in presence of obstacles 
different effects can be observed \cite{CPamm2017}. 
The barriers, depending 
on the system which is considered,
can either speed up or slow down the dynamics. 

For example, 
it is well known that the presence of obstacles can induce a 
sub--linear behavior with respect to time of the mean square distance 
traveled by particles undergoing Brownian motion. 
This phenomenon, called \emph{anomalous diffusion},
is observed in cells and in some cases it is explained as an effect 
due to the presence of macromolecules 
playing the role of obstacles for diffusing smaller molecules
\cite{Sbj1994,HFrpp2013,MHSbj2017,ESMCBjcp2014}.

In many other different contexts it has been proven
that the presence of an obstacle 
can surprisingly accelerate the dynamics.
In granular system
the out--coming flow, dramatically reduced
by the clogging at the exit, 
can be improved by placing an obstacle above the exit
\cite{TLPprl2001,ZGMPPpre2005,AMAGTOKpre2012,ZJGLAMprl2011}. 

A similar phenomenon is observed in pedestrian flows
\cite{Hrmp2001,BDsiamr2011,HMFBepBpd2001,HFMV2011,CMpA2013}
in case of panic, where
clogging at the door can be reduced by means of suitably 
positioned obstacles \cite{ABCKsjap2016,HFVn2000,HBJWts2005,EDLR2003}
that slow down pedestrian accumulation at the door
(the possibility of clustering far from 
the exit due to individual cooperation has been the object of study 
in \cite{CMpA2013,MCKBud2014,CMcrm2012}).
These unexpected phenomena are  
a sort of inverse Braess' paradox \cite{BNWts2005,Harfm2003}:
adding a road link to a road network can cause 
cars to take longer to cross the network, here, 
adding barriers results in a decrease 
of the time that particles need to cross a region of the space. 

These phenomena are discussed here in the very basic scenario of a 
symmetric random walk and it  is studied the effect of the barriers on the 
typical time, i.e., the \emph{residence time}, that a particle needs 
to cross a strip. 

In these terms the residence time issue has been posed in 
\cite{CKMSpre2016, CKMSSpA2016}, where the flow of particles 
entering an horizontal strip through the left end,
undergoing a random walk with exclusion inside the strip, and exiting it 
through the right end has been considered
\cite{FPvSpre2017}. 
In those papers a thorough study of the residence time properties as a function 
of the details of the dynamics, such as the horizontal drift, 
has been provided and in \cite{CKMSSpA2016} two different analytic 
tools have been developed. In \cite{CKMSpre2016}
it has been shown that, in some regimes, the residence time is not 
monotonic with respect to the size of the obstacle. 
This complex 
behavior has been related to 
the way in which particles are 
distributed along the strip at stationarity, more precisely, 
it has been explained in terms of the 
occupation number profile, which strongly 
depends on how particles interact due to the presence of the 
exclusion rule. 

Here, we consider the same geometry, but we assume that 
particles perform independent random walks in the strip. 
In other words we consider the average behavior of a single walker.  
Nevertheless, we observe surprising features of the system. 
We find that the residence time is non--monotonic with respect to the 
side lengths of the obstacle and the 
horizontal coordinate of its center. 
For suitable choices of the obstacle, the residence time in presence 
of the barrier is shorter than the one measured for the empty strip. 
We can say that placing a suitable obstacle
in the strip allows to select those particles that cross the strip 
in a shorter time. We also find that the same obstacle, placed in 
different positions along the strip, can either increase or decrease 
the residence time with respect to the empty strip case.
This complex behavior is not intuitive at all, 
indeed, it would be rather natural to infer that 
the presence of the obstacle 
increases the residence time since the channels 
flanking the obstacles are more difficult to be accessed 
by the particle. 

This problem has been studied in 
\cite{CCprep2017bol} in the framework of Kinetic Theory, 
more precisely for a model with particles moving according to the linear 
Boltzmann dynamics. 
Also in that case, it has been observed that the residence 
time is in some cases non--monotonic with respect to the 
geometrical parameters of the obstacle, such as its width and position. 

We can explain these phenomena as the  
consequence of the competition between two opposite effects.
The time that particles spend in the channels flanking 
the obstacle is smaller than the total time spent 
in the central part (the region containing the obstacle) 
of the strip in the empty case.
On the other hand, 
the time spent by the particles
in the regions of the strip on the left and on the 
right of the obstacle  is larger with respect to the empty case. 
These effects are due to the fact that 
it is more difficult for the walker 
to enter the central region of the strip, namely, 
one of the two channels formed by the obstacle.
The residence time behavior, hence, depends on which of the two effects
dominates the dynamics.

In this paper we also introduce a 1D model which mimics the 2D 
system. The presence of the obstacle is modelled via two defect sites, 
the left and the right one. The behavior of the particle sitting on 
one of these two special sites is similar to the one of the 2D particle 
moving in the columns adjacent to the obstacle. Indeed, 
we assume that the probability for the particle 
sitting on the left (resp.\ right) defect site to move to its 
right (resp.\ left) is smaller than $1/2$. 
The 1D model is studied both numerically and analytically, i.e., 
the residence time is computed exactly, even if we could not 
provide an explicit expression. The match between the numerical data and the 
analytic solution is perfect. 
The 1D model shows the same features as the 2D one and also the 
interpretation of the results is analogous. 
 
The paper is organized as follows. 
In Section~\ref{s:2D} we introduce the 2D model and discuss 
the related Monte Carlo results. 
In Section~\ref{s:1D} we propose the 1D analog and discuss both the 
numerical and the exact results. 
In Section~\ref{s:dimo} we prove the exact results. 
Finally, in Section~\ref{s:conclusioni} we summarize our conclusions.

\section{The 2D model}
\label{s:2D}
A particle performs a symmetric simple random walk on the 2D strip 
\begin{equation*}
 \Lambda = \left\{ x = (x_{1},x_{2}), x_{1} \in \{1,\ldots,L_{1}\}, x_{2} \in \{1,\ldots,L_{2}\} \right\}. 
\end{equation*}

The $1$ and the $2$ directions are respectively called 
\emph{horizontal} and \emph{vertical}.
The particle starts at a site in the first column on the left, namely, 
at a site $(1,x_2)$ with $x_2=1,\dots,L_2$
chosen at random with uniform probability. 
At each unit of time, the particle performs a move to one of the 
four neighboring sites with the same probability $1/4$. 
If the target site is in the horizontal boundary,
that is it belongs to the set  
$\{(x_{1},0), x_{1} \in \{1,\ldots,L_{1}\}\} 
\cup \{(x_{1},L_{2}+1), x_{1} \in \{1,\ldots,L_{1}\}\}$
the particle does not move, which means that 
the horizontal boundary is a reflecting surface.
If the target site belongs to the 
left or to the right vertical boundary 
$\{(0,x_{2}), x_{2} \in \{1,\ldots,L_{2}\}\} 
\cup \{(L_{1}+1,x_{2}), x_{2} \in \{1,\ldots,L_{2}\}\}$
the particle exits the system and the walk is stopped. 
Moreover, 
we shall consider a rectangular obstacle inside the strip, 
in the sense that, when one of the sites of this region 
will be chosen as target site for the move of the particle, the particle 
will not move. Thus, the sites in the obstacle are not accessible to 
the walker. The width and 
the height of the obstacle will be denoted respectively by 
$W$ and $H$.

The \emph{residence time} is defined as 
the mean time that the particle started at a uniformly chosen 
random site with abscissa $x_1=1$ takes 
to exit the strip through the right boundary. 
Sometimes, we shall address to the residence time as to the 
\emph{total} residence time to stress that it refers to the total 
time that the particle spends inside the strip. 
More precisely, one could 
consider the walk on the infinite strip 
$\mathbb{Z}\times\{1,\dots,L_2\}$ and define the residence time as 
the mean of the first hitting time for a particle started at a site 
$(1,x_2)$, with $x_2=1,\dots,L_2$ chosen at random with uniform 
probability, to the set of 
sites with $x_1=L_1+1$ conditioned 
to the event that the particle reaches such a subset before 
visiting the set of sites with abscissa $x_1=0$. 

We shall compute the residence time 
by simulating many particles and averaging 
the time that each of them needs to exit, paying attention to the fact that 
only those particles which effectively exit through 
the right boundary will contribute to the average, whereas those 
exiting through the left boundary will be discarded. 

As in the case discussed in \cite{CCprep2017bol} 
in the framework of Kinetic Theory, 
we find a surprising result: the residence time is not monotonic 
with respect to the geometrical parameters of the obstacle, 
such as its position and its size. 
We show, also, that obstacles can increase or decrease the residence time with 
respect to the empty strip case depending on their side lengths 
and on their position. In some cases, 
one of these parameters controls 
a transition from the increasing to the decreasing effect.
We stress that in some cases the residence time measured in presence 
of an obstacle is smaller than the one measured for the empty strip, 
that is to say, the obstacle is able to select those 
particles that cross the strip faster.

We now discuss our results for different choices of the 
obstacle and postpone our interpretation to the 
end of this section. 
All the details about the numerical simulations  
are in the figure captions. 
The statistical error, since negligible, 
is not reported in the picture. 

\begin{figure}[t]
\vspace{-2.5 cm}
\centerline{%
\hspace{3.7 cm}
{\includegraphics[width=.9\textwidth]{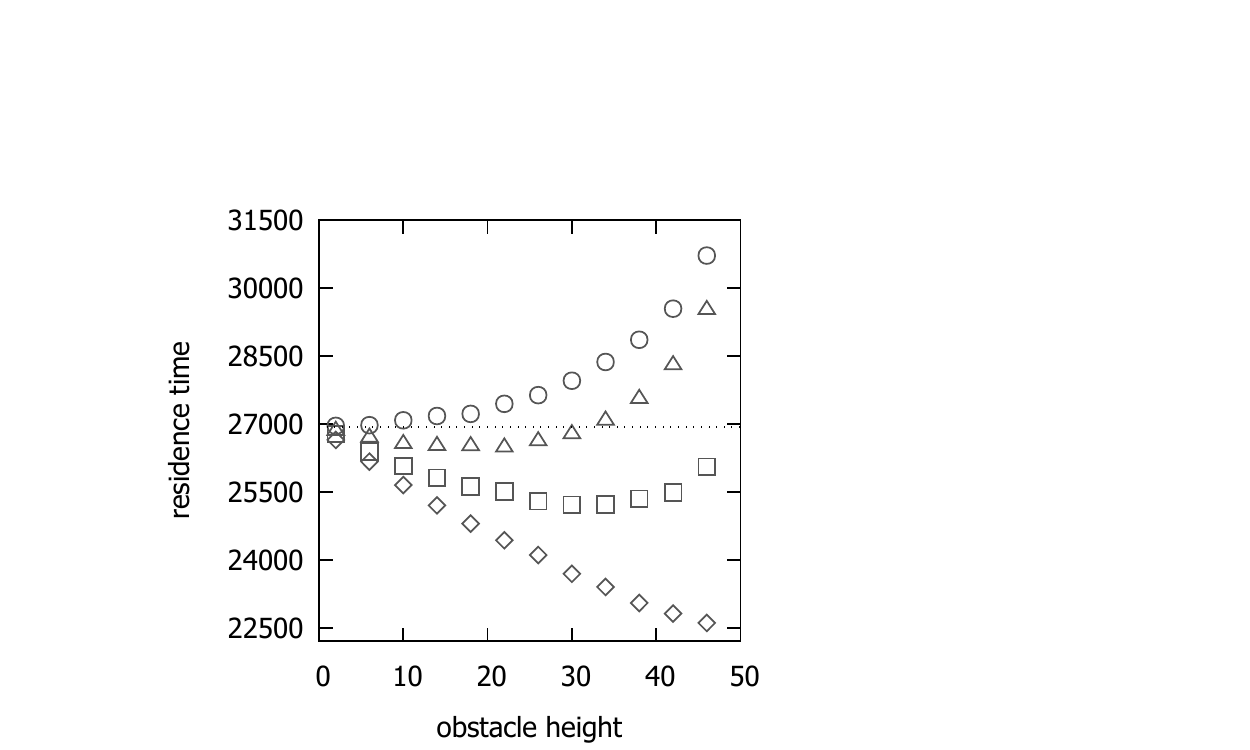}}
}
\caption{Residence time vs.\ obstacle height. 
The obstacle is placed at the center of the strip and its 
 width is 
$W=2$ (disks).
$W=20$ (triangles), 
$W=40$ (squares), and 
$W=60$ (diamonds).
Simulation parameters:
$L_1=200$,
$L_2=50$,
total number of inserted particles $5 \cdot 10^7$.
The total number of particles exiting through the right boundary
decreases when the obstacle height is increased
from 
$2.49 \cdot 10^{5}$ (empty strip) to $1.69\cdot 10^{5}$ (disks),
$0.99\cdot 10^{5}$ (triangles), and 
$0.68\cdot 10^{5}$ (squares) for $H=46$.
The dashed line at $26930$ represents the value of the residence time measured
for the empty strip.
}
\label{f:c1_4_rw2}
\end{figure}

In Figure~\ref{f:c1_4_rw2} 
we report the residence time as a function of the 
obstacle height. The obstacle is placed at the center of the strip
and its width is $W=2$ (disks), 
$W=20$ (triangles), 
$W=40$ (squares), and 
$W=60$ (diamonds).
In the case of a thin barrier, starting from the empty strip value, 
the residence time increases with the height of the obstacle. 
For a wider obstacle, an 
a priori not intuitive result is found: 
the dependence 
of the residence time on the obstacle height is not monotonic. 
In the case $W=20$, 
starting from the empty strip value, the residence time 
decreases up to height $20$ and then increases to values above the 
empty strip one. 
This effect is even stronger if the width 
of the obstacle is further increased. 

\begin{figure}[!h]
\vspace{-2.5 cm}
\centerline{%
\hspace{3.7 cm}
{\includegraphics[width=.9\textwidth]{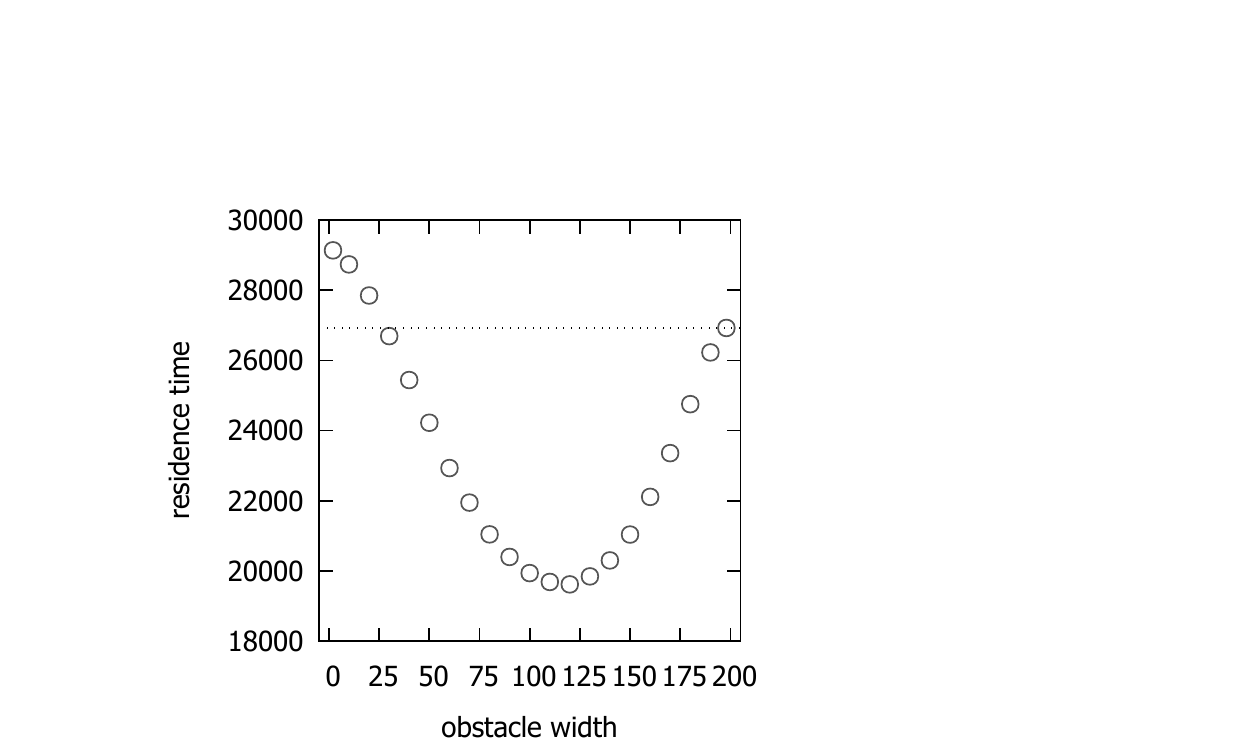}}
}
\caption{Residence time vs.\ obstacle width.
The obstacle is placed at the center of the strip and its 
 height is $H=40$.
The total number of particles exiting through the right boundary
decreases when the obstacle width is increased from 
$2 \cdot 10^{5}$ for $W=2$ to $0.5\cdot 10^{5}$ for $W=198$.
The parameters of the simulation and 
the dashed line are as in Figure~\ref{f:c1_4_rw2}.
}
\label{f:c5_rw2}
\end{figure}

In Figure~\ref{f:c5_rw2} we plot the residence time as a 
function of the obstacle width. 
The obstacle is placed at the center of the strip
and its height is $H=40$. 
When the barrier is thin the residence time is larger than 
the one measured in the empty strip case, but, when the width 
is increased, the residence time decreases and at about $26$ it 
becomes smaller than the empty case value. 
The minimum is reached at about $120$ (recall 
that the length of the strip is $L_{1}=200$ 
in this simulation), then the residence time 
increases to the empty strip value 
which is reached when the obstacle is as long as the entire strip.
This is intuitively obvious, since in such a case 
the lattice consists of 
two independent channels having the same length as the original strip. 

\begin{figure}[t]
\vspace{-2.5 cm}
\centerline{%
\hspace{3.7 cm}
{\includegraphics[width=.9\textwidth]{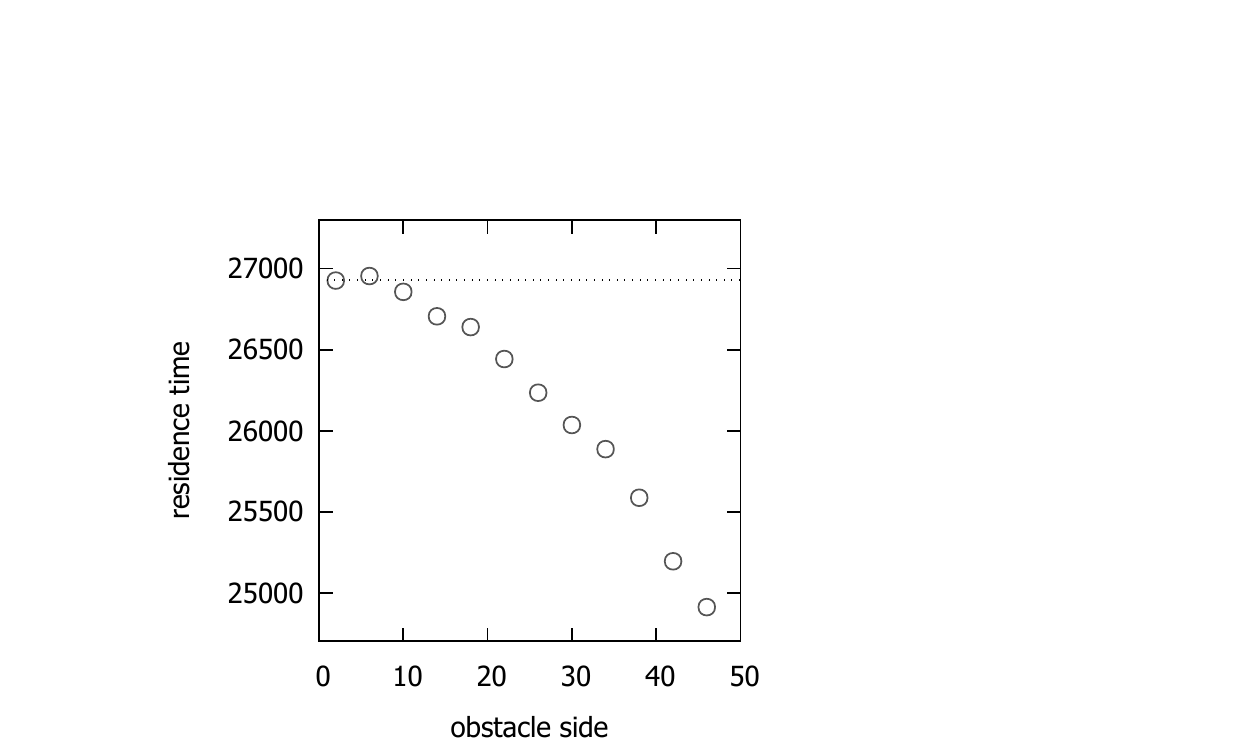}}
}
\caption{Residence time vs.\ squared obstacle side length.
The squared obstacle is placed in the middle of the strip.
The total number of particles exiting through the right boundary
decreases when the obstacle side length is increased from 
$2.49 \cdot 10^{5}$ (empty strip) to $0.63\cdot 10^{5}$ for side length equal to $46$.
The parameters of the simulation and 
the dashed line are as in Figure~\ref{f:c1_4_rw2}.
}
\label{f:c6_rw2}
\end{figure}

In Figure~\ref{f:c6_rw2} a centered square obstacle is 
considered. The residence time as a function of its side length is 
reported. Although small oscillations, reasonably due to numerical 
approximations, 
are visible, the behavior appears to be monotonically 
decreasing. 

\begin{figure}[t]
\vspace{-2.5 cm}
\centerline{%
\hspace{3.7 cm}
{\includegraphics[width=.9\textwidth]{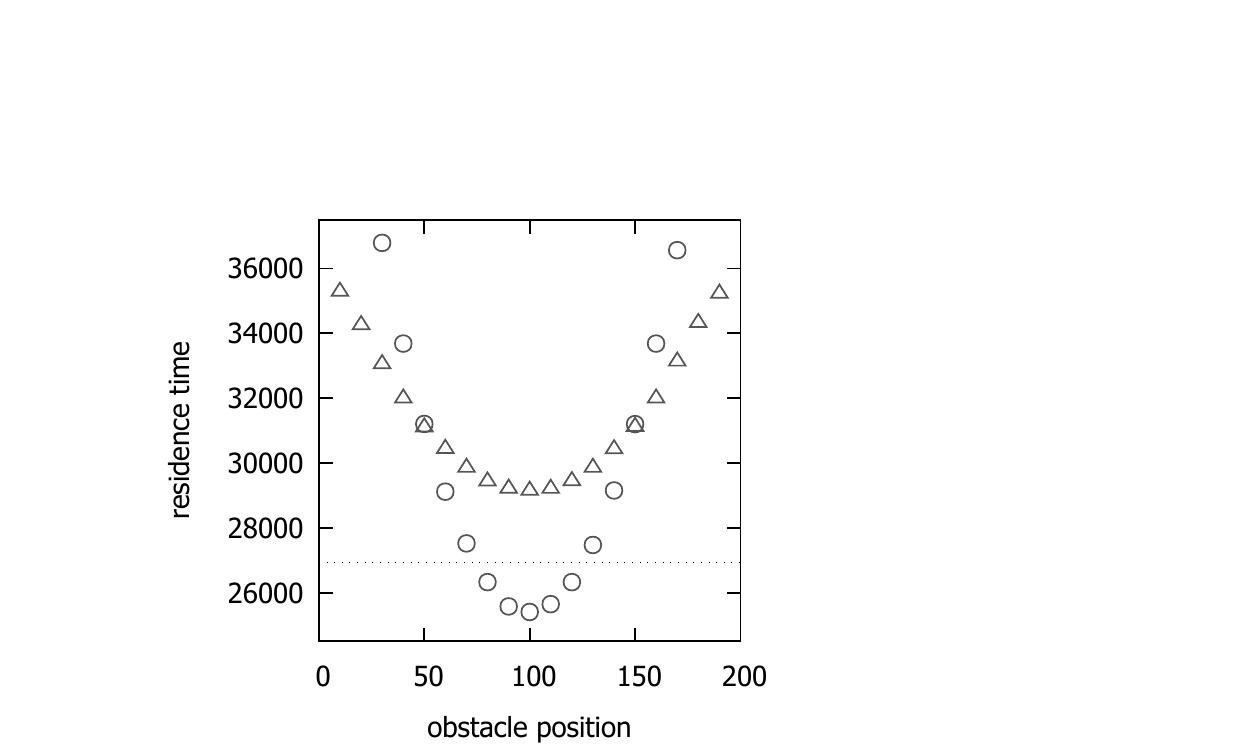}}
}
\caption{Residence time
vs.\ position of the center of the obstacle.
Disks refer to a squared obstacle with side length $40$, 
whereas triangles refers to a rectangular obstacle 
with width $W=2$ and height $H=40$.
The total number of particles exiting through the right boundary
is approximately $1.24\cdot 10^5$ (disks) and 
$2.01\cdot 10^5$ (triangles) and depends poorly on the 
obstacle position. 
The parameters of the simulation and 
the dashed line are as in Figure~\ref{f:c1_4_rw2}.
}
\label{f:c7_8_rw2}
\end{figure}

Finally, 
in Figure~\ref{f:c7_8_rw2} we show that, and this is really surprising,
the residence time is not monotonic even
as a function of the position of the center of the obstacle. 
Disks refer to a squared obstacle of side length $40$, 
whereas triangles refer to a thin rectangular obstacle with 
width $W=2$ and height $H=40$. 
In both cases the residence time is not monotonic and attains its 
minimum value when the obstacle is placed in the center of the strip. 
In the squared obstacle case, when the abscissa of the center of the obstacle lies between 
$75$ and $125$ the residence time in 
presence of the obstacles is smaller than the corresponding value for 
the empty strip. On the other hand, for the thin rectangular 
obstacle, even if the non--monotonic behavior is found, 
the residence time is always larger than in the empty strip case. 
This fact is consistent with the results plotted in 
Figure~\ref{f:c1_4_rw2}.

The results that we found in the numerical experiments reported in 
Figures~\ref{f:c1_4_rw2}--\ref{f:c7_8_rw2} can be summarized as follows: 
the residence time strongly depends on the obstacle geometry and position. 
In particular it seems that large centered obstacles favor the selection 
of particles crossing the strip faster than in the empty strip case. 

In order to explain our observations, 
following \cite{CCprep2017bol}, we partition the strip into 
three parts: the rectangular region on the left of the obstacle, 
the rectangular region on the right of the obstacle and the 
remaining central part containing the obstacle.
As we will see later, 
the residence time behavior 
is consequence of two effects in competition: 
the total time spent by the particles in the channels between the
obstacle and the horizontal boundary is smaller than the total time spent 
in the central part of the strip in the empty case.
On the contrary,
the total time spent both in the left and in the right part of the strip 
is larger with respect to the empty case. 
Both these two effects can be explained remarking that, when the 
obstacle is present, 
it is more difficult for the walker 
to enter the central region of the strip, namely, 
one of the channels flanking the obstacle.
The total residence time trend depends on which of the two effects
dominates the dynamics of the walker.

\begin{figure}[!h]
\vspace{-1. cm}
\centerline{%
\hspace{2.5 cm}
{\includegraphics[width=.7\textwidth]{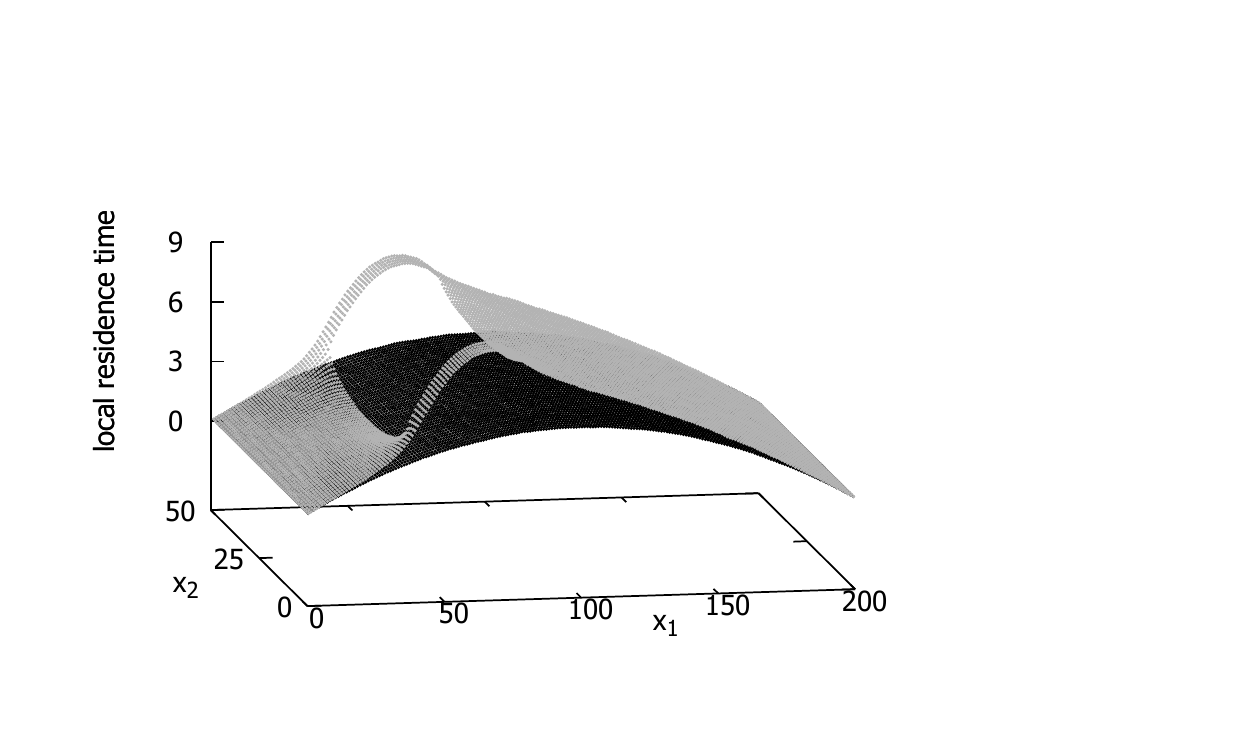}}
}
\vspace{-1. cm}
\caption{Mean time spent by the walker crossing the strip 
in each site of the strip (local residence time) for the empty strip case 
(black) and in presence of the obstacle (gray).
Data are those of the experiment described by the disks 
in Figure~\ref{f:c7_8_rw2}.
The obstacle is a squared obstacles with side length $40$ 
placed at the site with abscissa $60$.
}
\label{f:int01}
\end{figure}

To illustrate our interpretation of the phenomenon we describe in 
detail the walker behavior referring to the experiment 
associated with the disks in Figure~\ref{f:c7_8_rw2}.
In Figure~\ref{f:int01} we plot the 
mean time spent by the walker crossing the strip
in each site of the strip. This quantity will be addressed as the 
\emph{local residence time}. 
The gray surface in the picture refers to the obstacle case, whereas the 
black surface is related to the empty strip case. The data in the 
picture have been collected in the case in which the center 
of a squared obstacle with side length $40$ is placed at the site 
with coordinates $(80,25)$.
The graph shows that in average in each site of the strip the particle spends 
a time larger than the time it spends at the same site in the 
empty strip case. This seems to be in contrast with the fact that 
the (total) residence time in the strip 
can be smaller when the obstacle is present. 
Indeed, this can happen since the sites of the strip 
falling in the obstacle region are never visited by the walker. 
It can then happen that the sum of the local residence times 
associated with sites in the central part of the strip in presence of the 
obstacle is smaller than
the same sum computed in the empty strip case. 

\begin{figure}[t]
\vspace{-2.5 cm}
\centerline{%
\hspace{3.7 cm}
{\includegraphics[width=.9\textwidth]{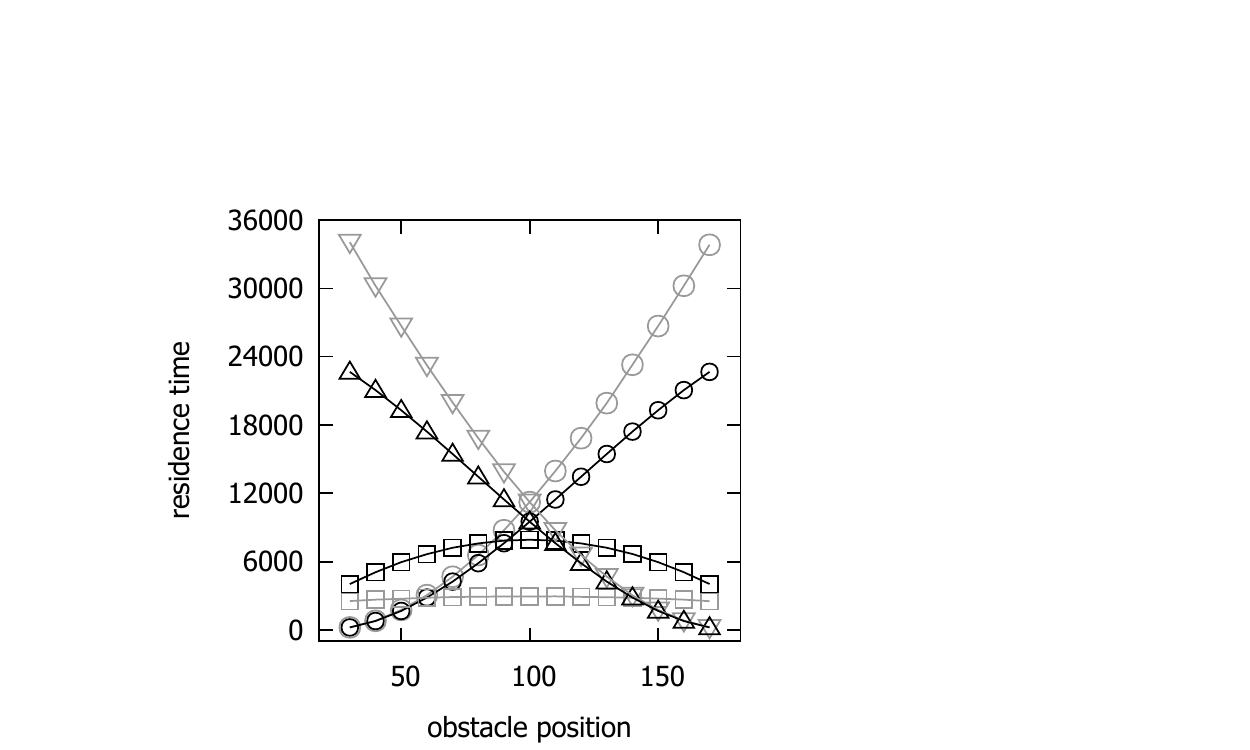}}
}
\caption{Total residence time in the left (circles), central (squares), 
and right (triangles) part of the strip
in presence of the obstacle (gray) and in the empty strip case (black). 
The experiment associated with the disks in Figure~\ref{f:c7_8_rw2}
is considered.
}
\label{f:int02}
\end{figure}

Results in Figure~\ref{f:int01} can be interpreted as follows.
The local residence times in the left and in the right regions are larger 
with respect to the empty case since for the particle it is more difficult
to access the central region and, thus, it will spend more time in the 
lateral parts of the strip. On the other hand, once the particle enters 
into one of the two central channels, it will take in average the 
same time to get back to one of the two lateral parts of the strip that 
it would take in absence of the obstacle. But, since, the number of the 
available sites in the central part is smaller when the obstacle is present, 
the local residence time will be larger. 

The Figure~\ref{f:int01} gives some new insight in the motion of the walker, 
but it is not sufficient to explain the residence time 
behavior discussed above. In order to get some insight into this, we
compute the respective times 
spent by the particle in the left, central and right 
region of the strip. This is done in Figure~\ref{f:int02}, where 
data referring to the experiment associated with the disks in 
Figure~\ref{f:c7_8_rw2} are reported. 
First, one should note that
the total residence time in the left 
and in the right part of the strip are increased when the obstacle 
is present, this is due to the fact that for the particle it is 
more difficult to enter the central part when 
the obstacle is present. 
Moreover, precisely for the same reason the trajectory of the walker 
from its starting point to its exit from the strip
will visit the channels in the central region of the strip 
a number of time smaller than the number of times that the particle
visits the central region of the strip in the empty strip case. 
Thus, the residence time in the central part of the strip results 
to be smaller when the obstacle is present. 

Hence, the behavior of the (total) residence time data reported as disks in 
Figure~\ref{f:c7_8_rw2}
can be explained as follows: if the center of the obstacle 
is close to the left boundary (say its abscissa is smaller than $75$)
the effect in the right region of the strip dominates the one in the central 
region and the (total) residence time is increased (the effect in the 
left region in this case is negligible). 
On the other hand, 
if the center of the obstacle 
is close to the center of the strip (say its abscissa is between $75$
and $125$)
the effect in the central region dominates and the (total) residence 
time is decreased. 
Finally, 
if the center of the obstacle 
is close to the right boundary (say its abscissa is larger than $125$)
the effect in the left region of the strip dominates the one in the central 
region and the (total) residence time is increased (the effect in the 
right region in this case is negligible). 

The behavior of the residence time in connection with all the experiments 
illustrated in Figures~\ref{f:c1_4_rw2}--\ref{f:c7_8_rw2} can be 
explained similarly. 

\section{The 1D model}
\label{s:1D} 
In this section we propose 
a one--dimensional reduction of the problem based on a symmetric 
simple random walk with two defect sites. 
We actually prove that the behaviors of the 1D system are similar to those  
discussed above
and that the Monte Carlo data are  fully supported by 
exact analytical computations.

We consider a simple random walk on $\{0,1,\dots,L\}$.
The sites $0$ and $L$ are absorbing, 
so that when the particle reaches one of these two sites the walk is stopped. 
All the sites $1,\dots,L-1$ are \emph{regular} excepted for two 
sites called \emph{defect} or \emph{special} sites. 
The \emph{first} or \emph{left} defect site is the site $n+1$ and 
the \emph{second} or \emph{right} defect site is the site $n+h+2$, 
with $n=1,2,\dots,L-5$ and $h=1,2,\dots,L-n-4$.
The parameters $n$ and $h$ are chosen in such a way that the left 
defect site cannot be $1$, the right defect site cannot be $L-1$, and 
there is at least one regular site separating the two defect sites. 
The number of regular sites on the left of the left defect site is $n$
and
the number of regular sites in the region between the two defect 
sites is $h$. We let $w=L-(n+h+3)$ be the number of regular 
sites on the right of the right defect site. 

At each unit of time the walker 
jumps to a neighbouring site according to the following rule:
if it is on a regular site, then it 
performs a simple symmetric random walk.
If it is at the left defect site it jumps with 
probability $\lambda$ to the right, with probability $1-\lambda-\epsilon$ 
to the left, and with probability $\epsilon$ it does not move.  
If it is at the right defect site it jumps with 
probability $\lambda$ to the left, with probability $1-\lambda-\epsilon$ 
to the right, and with probability $\epsilon$ it does not move.  
Here, $\lambda\in(0,1)$ and $\epsilon\in[0,1)$.

The array $1,\dots,L-1$ will 
be called the \emph{lane}. The sites $0$ and $L$ will be, 
respectively, called the \emph{left} and \emph{right exit} of the lane. 

This 1D model is a toy model for the 2D system that we have 
discussed in Section~\ref{s:2D}. Indeed, the left defect 
site $n+1$ mimics the sites in 
the first column of the 2D strip on the left of the obstacle: 
the 2D walker in such a column has a probability to move to the right
smaller than the probability to move to the left. Similarly, 
the right defect site $n+h+1$ mimics the sites in the first column 
to the right of the obstacle. Let us stress that 
the sites $n+1+1,\dots,n+h$ are regular, since when the 2D walker 
enters one of the two channels flanking the obstacle its probability 
to move to the right is equal to that to move to the left. 

In this framework the residence time is defined by starting 
the walk at site $1$ and computing 
the typical time that the particle takes to reach the site $L$ provided 
the walker reaches $L$ before $0$.
More precisely, we let $x_t$ be the position of the walker at time $t$ and 
denote by $\mathbb{P}_k$ and $\mathbb{E}_k$ the probability associated to the 
trajectories of the walk and the related average operator 
for the walk started at $x_0=k$ with $k=1,\dots,L-1$. 
We let
\begin{equation}
\label{one000}
T_i=\inf\{t>0,\,x_t=i\}
\end{equation}
be the \emph{first hitting time to} $i$, with the convention that 
$T_i=\infty$ if the set $\{t>0,\,x_t=i\}$ is empty, i.e., the 
trajectory does not reach the site $i$. 
The main quantity of interest is the \emph{residence time} or
\emph{total residence time} 
\begin{equation}
\label{one010}
R
=
\mathbb{E}_1[T_L|T_L<T_0]
=
\sum_{t=1}^\infty
t\mathbb{P}_1[T_L=t|T_L<T_0]
\;\;.
\end{equation}
Note that the residence time is defined for the walk started at $x_0=1$ and 
the average is computed conditioning to the event $T_L<T_0$,  namely,
conditioning to the fact that the particle exits the lane through the 
right exit. 

As in the 2D case discussed in Section~\ref{s:2D}, 
we shall compute numerically the residence time 
by simulating many particles and averaging 
the time that each of them takes to exit through the right 
ending point, discarding all the particles 
exiting through the left ending point. 
But we stress that in this 1D model it is also possible to compute 
exactly the residence time. In this section we shall discuss our 
findings and in each plot the solid lines will represent the exact 
result which will be discussed in the following Section~\ref{s:dimo}.

We now discuss our results for different choices of the 
parameter which are the analog of the cases 
considered in Section~\ref{s:2D} for the 2D model. 
All the details about the numerical simulations  
are in the figure captions. The statistical error, since negligible, 
is not reported in the picture. 
We carry out the simulations with the following choice of
the parameters:
\begin{equation}
\label{one020}
\epsilon=\frac{1}{2}p
\;\;\textrm{ and }\;\;
\lambda=\frac{1}{2}(1-p)
\end{equation}
with $p\in[0,1)$, so that $\epsilon\in[0,1/2)$ and 
$\lambda\in(0,1/2]$. Note that with such a choice the probability 
to move left (resp.\ right) for the particle sitting 
at the left (resp.\ right) defect site is 
$1-\lambda-\epsilon=1/2$. 
Note that for $p$ equal zero we recover the symmetric simple random walk, 
which mimics the 2D empty strip.

\begin{figure}[t]
\vspace{-2.5 cm}
\centerline{%
\hspace{3.7 cm}
{\includegraphics[width=.9\textwidth]{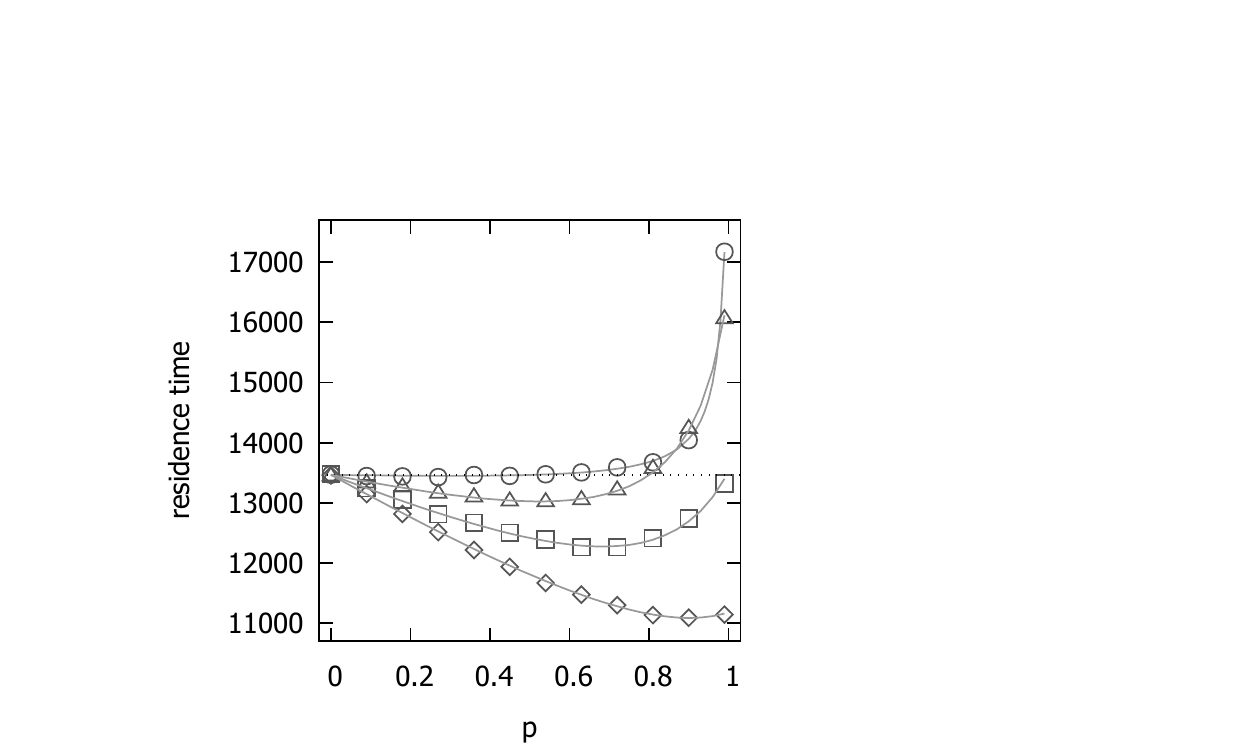}}
}
\caption{Residence time vs.\ $p$.
Simulation parameters:
total number of inserted particles $5 \cdot 10^7$,
$L=201$,
$n=98$ and $h=2$ (disks),
$n=89$ and $h=20$ (triangles),
$n=79$ and $h=40$ (squares), 
and
$n=69$ and $h=60$ (diamonds).
The total number of particles exiting through the right exit
decreases when $p$ grows from 
$2.48\cdot 10^{5}$ (no defect is present, namely, $p=0$)
to 
$1 \cdot 10^{5}$ (disks), 
$0.22 \cdot 10^{5}$ (triangles), 
$0.12 \cdot 10^{5}$ (squares), 
and
$0.08 \cdot 10^{5}$ (diamonds) for $p=0.99$.
The dashed line at $13466$ line represents the value of the residence time 
measured in absence of defect sites ($p=0$).
The solid line is the exact solution. 
}
\label{f:c1_4_rw1}
\end{figure}

The case reported in Figure~\ref{f:c1_4_rw1} is the analog of the 
case discussed in Figure~\ref{f:c1_4_rw2} in the 2D setting. 
Indeed, the residence time is plotted as a function of the parameter $p$ 
increasing from $0$ to $0.99$ and this mimics the increase of 
the height of the obstacle considered in Figure~\ref{f:c1_4_rw2}.
Moreover, the two defect sites are symmetric with respect to 
the middle point of the lane and the number of regular sites 
between them is chosen equal to $2$, $20$, $40$, and $60$ mimicking 
the different obstacle widths considered in Figure~\ref{f:c1_4_rw2}.
The data show a behavior similar to that reported in Figure~\ref{f:c1_4_rw2}
in the 2D case:
in the case $h=2$ (the defect sites are close to each other) 
the residence time increases with $p$. 
For a wider obstacle, the non--monotonic behavior 
is recovered. 
In the case $h=20$, 
starting from the empty strip value, the residence time 
decreases up to $p\sim0.55$ and then it increases to values above the 
$p=0$ case.
This effect is even stronger if $p$ is further increased. 

\begin{figure}[t]
\vspace{-2.5 cm}
\centerline{%
\hspace{3.7 cm}
{\includegraphics[width=.9\textwidth]{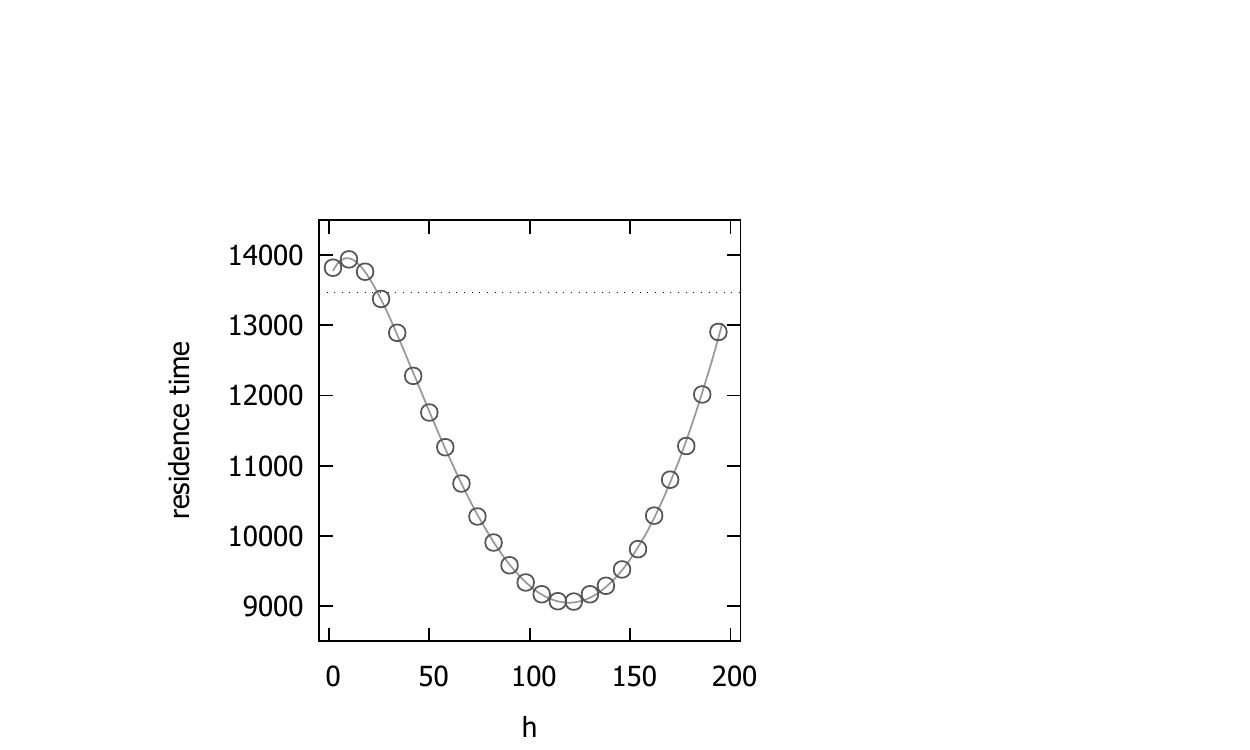}}
}
\caption{Residence time vs.\ $h$ (even)
for $p=0.84$, $L=201$, $n=(L-h-3)/2$,
and 
total number of inserted particles $5 \cdot 10^7$.
The dashed and the solid lines are as in Figure~\ref{f:c1_4_rw1}.
The total number of particles exiting through the right exit
decreases when $h$ is increased 
from $2.31 \cdot 10^{5}$ for $h=2$ to $0.41\cdot 10^{5}$ for $h=194$.
}
\label{f:c5_rw1}
\end{figure}

The case reported in Figure~\ref{f:c5_rw1} is the analog of the 
case discussed in Figure~\ref{f:c5_rw2} in the 2D setting. 
Indeed, the residence time is plotted as a function of the parameter $h$ 
increasing from $2$ to $198$ with the two defect sites 
symmetric with respect to the middle point of the lane. 
This case mimics the increase of the width of the centered rectangular 
obstacle reported in Figure~\ref{f:c5_rw2}.
When $h$ is small the residence time is larger than 
the one measured for $p=0$, but, when $h$ 
is increased, the residence time decreases and at about $25$ it 
becomes smaller than the $p=0$ case.
The minimum is reached at about $120$ (recall the lane is long $201$ sites 
in this simulation), then the residence time 
increases towards the $p=0$ value. 

\begin{figure}[t]
\vspace{-2.5 cm}
\centerline{%
\hspace{3.7 cm}
{\includegraphics[width=.9\textwidth]{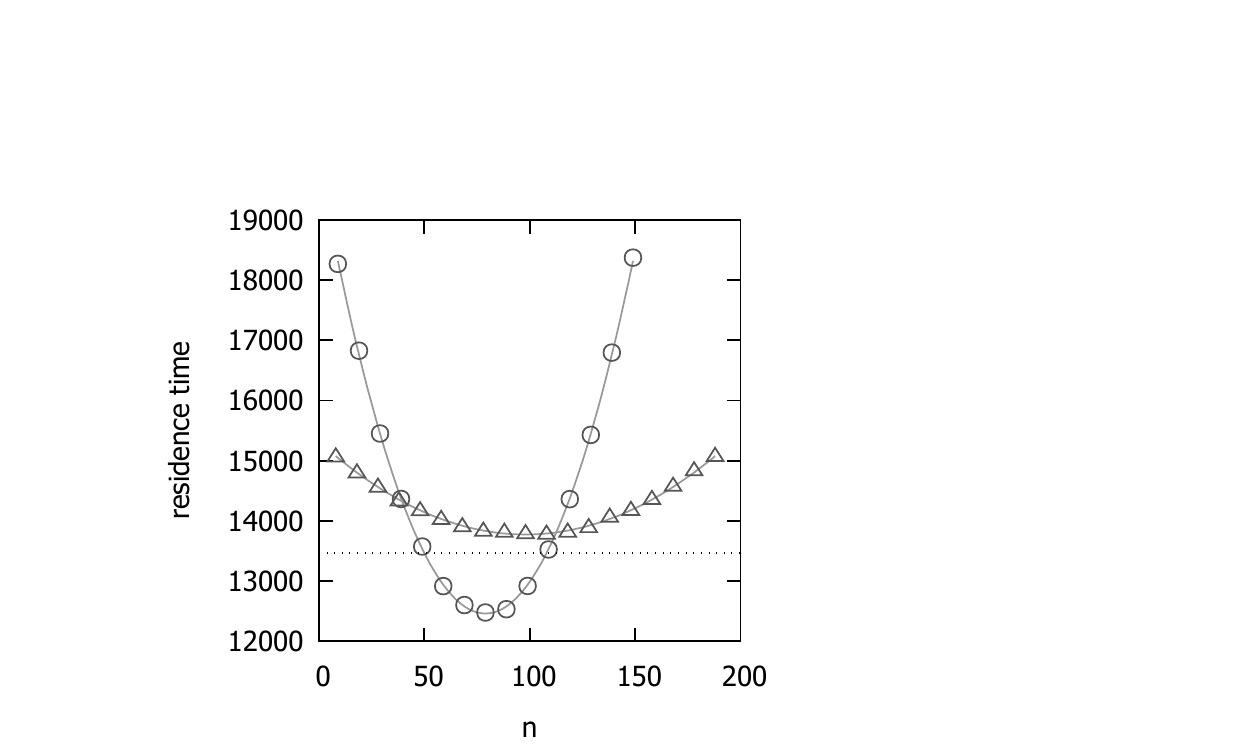}}
}
\caption{Residence time vs.\ $n$ 
for $p=0.84$, $L=201$, $h=40$ (disks), $h=2$ (triangles),
and 
total number of inserted particles $5 \cdot 10^7$.
The dashed and the solid lines are as in Figure~\ref{f:c1_4_rw1}.
The total number of particles exiting through the right exit
is approximately equal to 
$1.2 \cdot 10^{5}$ (disks) and $2.3\cdot 10^{5}$ (triangles).
}
\label{f:c7_8_rw1}
\end{figure}

In this 1D setting it is not really clear how to construct an analog 
for the experiment in Figure~\ref{f:c6_rw2}, where a squared 
centered obstacle was considered. 
On the other hand, 
the case reported in Figure~\ref{f:c7_8_rw1} is the analog of the 
case discussed in Figure~\ref{f:c7_8_rw2} in the 2D setting. 
Indeed, the residence time is plotted as a function of the parameter $n$
in the two cases $h=40$ (disks) and $h=2$ (triangles).
This case mimics the increase of the abscissa of the 
center of the obstacle reported in Figure~\ref{f:c7_8_rw2}.
In both cases the residence time is non--monotonic and attains its 
minimum value when the defect sites are symmetric with respect to 
the center of the lane. 
In the $h=40$ case, when $n$ lies approximately between 
$50$ and $110$ the residence time 
is smaller than the corresponding value for 
the case $p=0$. 
On the other hand, for $h=2$,
even if the non--monotonic behavior is recovered, 
the residence time is always larger than the one measured in the $p=0$ case. 
This fact is consistent with the results plotted in 
Figure~\ref{f:c1_4_rw1}.

\begin{figure}[t]
\vspace{-2.5 cm}
\centerline{%
\hspace{3.7 cm}
{\includegraphics[width=.9\textwidth]{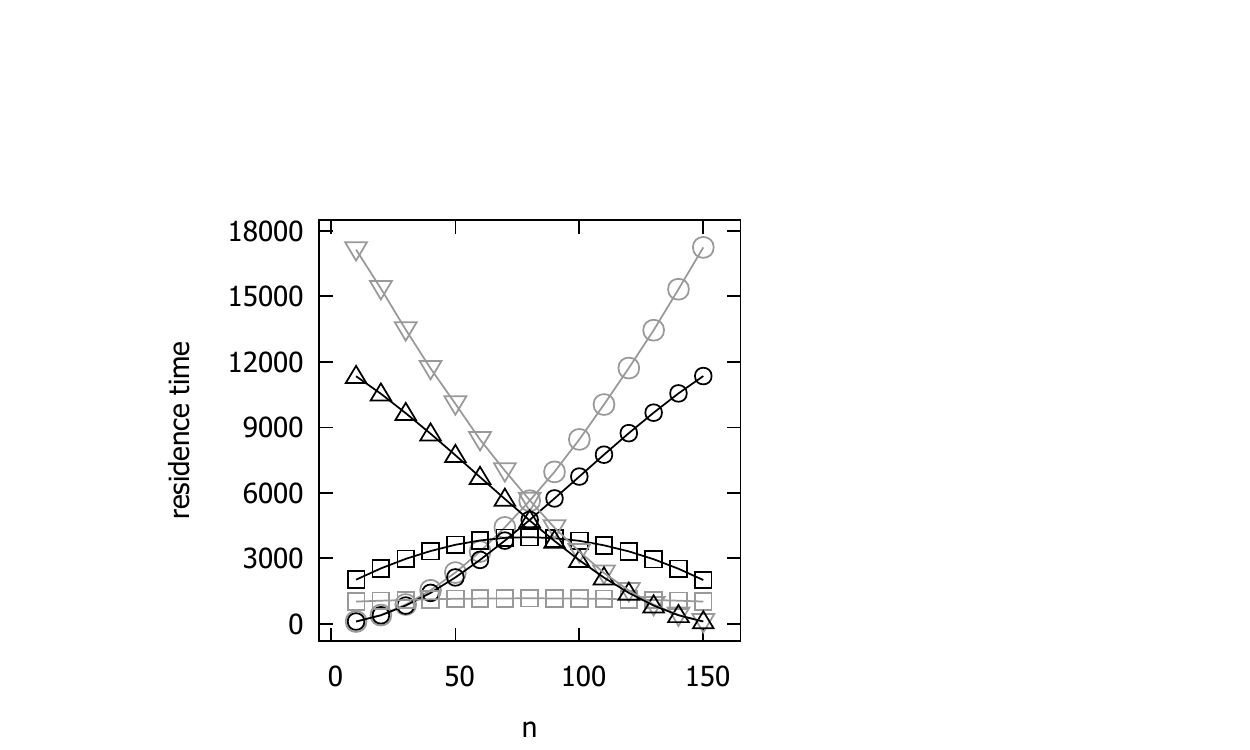}}
}
\caption{Total residence time in the left (circles), central (squares), 
and right (triangles) region of the lane
in presence of the defect sites (gray) and in the $p=0$ case (black). 
The experiment associated with the disks in Figure~\ref{f:c7_8_rw1}
is considered.
}
\label{f:int02-rw1}
\end{figure}

In order to explain these observations, 
similarly to what we did in the 2D case 
and in \cite{CCprep2017bol}, we partition the lane into 
three parts: 
the part of the lane on the left of the left defect (left region), 
the part of the lane between the two defect sites (central region), 
and
the part of the lane on the right of the right defect (right region).
As in the 2D case,
the residence time behavior 
is consequence of two effects in competition: 
the total time spent by the particles in the central region 
is smaller than the total time spent in the same region in 
absence of defect sites ($p=0$).
On the contrary,
the total time spent both in the left and in the right region 
is larger with respect to the time spent there in the $p=0$ case. 
Both these two effects can be explained remarking that, in presence 
of defect sites, it is more difficult for the walker 
to enter the central region of the lane.
The total residence time trend depends on which of the two effects
dominates the dynamics of the walker.

These remarks are illustrated in Figure~\ref{f:int02-rw1},
data referring to the experiment associated with the disks in 
Figure~\ref{f:c7_8_rw1} are reported. 
Again, one notes that
the total residence time in the left 
and in the right regions of the lane are increased when the defect sites 
are present, this is due to the fact that for the particle it is 
more difficult to enter the central region in such a case. 
Moreover, precisely for the same reason the trajectory of the walker 
from its starting point $1$ to its exit from the lane
will visit the central region of the lane
a number of time smaller than the number of times that the particle
visits such a region in the $p=0$ case. 
Thus, the residence time in the central region results 
to be smaller in presence of the defect site.
Finally, similarly to what we did in the 2D case, the results in 
Figure~\ref{f:int02-rw1} allows a complete interpretation of the 
residence time behavior depicted by the disks in Figure~\ref{f:c7_8_rw1}
(note that the maximum value of the variable $n$ for the disks in 
Figure~\ref{f:c7_8_rw1} is $150$).

\section{Analytic results}
\label{s:dimo} 
\begin{figure}[t]
\vspace{0.5 cm}
\centerline{%
\hspace{-0.5 cm}
{\includegraphics[width=.3\textwidth]{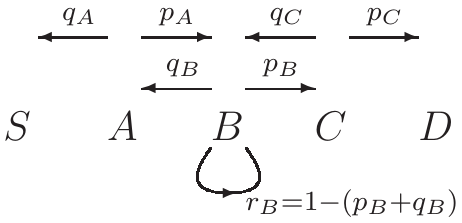}}
}
\caption{Schematic representation of the five state chain model.}
\label{f:5stati}
\end{figure}

In this section we derive exact, though not explicit, expressions for the 
residence time defined in Section~\ref{s:1D}. To compute the 
residence time, we shall make 
use of the following result on a five state chain: the states 
are $S$, $A$, $B$, $C$, and $D$. The jump probabilities are as 
depicted in the figure~\ref{f:5stati} and the chain is started at time 
$0$ in $B$. 
We prove that the probability $Q_k$, with $k\ge1$, for the 
chain to reach $D$ before $S$ and 
return $k-1$ times  
to the site $B$ before reaching $D$ is 
\begin{equation}
\label{dimo000}
Q_k
=
p_Bp_C[r_B+q_Bp_A+p_Bq_C]^{k-1}
\;\;,
\end{equation}
where $r_B=1-(p_B+q_B)$.
Indeed, 
\begin{displaymath}
\begin{split}
Q_k
&
=
p_Bp_C 
\sum_{r=0}^{k-1} 
{{k-1}\choose{r}} (p_Bq_C)^{k-1-r}
\\
& 
\phantom{mmmmmmmm}
\times 
\sum_{s=0}^r
{{r}\choose{s}} 
 (q_Bp_A)^s (r_B)^{r-s}
\end{split}
\end{displaymath}
where $r$ counts the number of times that, starting from $B$, 
the chain either jumps to $A$ or it stays in $B$ and 
$s$ counts the number of times that starting from $B$ it jumps to $A$. 
The equation \eqref{dimo000} is then proven by 
using the binomial theorem.

We now consider again to the 1D walk defined in Section~\ref{s:1D}.
To compute the residence time we introduce the \emph{local times}, 
i.e., the time spent by a trajectory at site $i$ defined as 
\begin{equation}
\label{dimo010}
\tau_i
=|\{t>0,x_t=i\}|
\end{equation}
for any $i=1,\dots, L-1$, where 
$|A|$ denotes the cardinality of the set $A$. 
Provided $T_L$ is finite, 
we have that
\begin{equation}
\label{dimo020}
T_L
=
\sum_{i=1}^{L-1}\tau_i
\;\;.
\end{equation}
Hence the residence time $R$ defined in \eqref{one010} 
can be expressed as 
\begin{equation}
\label{dimo030}
R=
\sum_{i=1}^{L-1}\mathbb{E}_1[\tau_i|T_L<T_0]
\end{equation}
and for all $i \in \{1,\ldots,L-1\}$
\begin{equation}
\label{dimo040}
\begin{split}
&\mathbb{E}_1[\tau_i|T_L<T_0]
\vphantom{\bigg\{}
\\
&
\phantom{m}
=\frac{\mathbb{P}_1[T_i<T_0]}{\mathbb{P}_1[T_L<T_0]}
\frac{p_Bp_C}{[1-(r_B+q_Bp_A+p_Bq_C)]^2}
\;\;,
\end{split}
\end{equation}
where we defined the quantities
\begin{equation}
\label{dimo050}
\begin{array}{ll}
p_A=\mathbb{P}_{i-1}(T_i<T_0),
&
q_A=\mathbb{P}_{i-1}(T_0<T_i),\\
p_B=\mathbb{P}_i(x_1=i+1),
&
q_B=\mathbb{P}_i(x_1=i-1),\\
p_C=\mathbb{P}_{i+1}(T_L<T_i),
&
q_C=\mathbb{P}_{i+1}(T_i<T_L).
\end{array}
\end{equation}

Note that $p_A+q_A=1$, $p_{B} + q_{B} + r_{B} = 1$, and $p_C+q_C=1$.
Indeed, we have
\begin{displaymath}
\mathbb{E}_1[\tau_i|T_L<T_0]
=
\sum_{k=1}^\infty k \mathbb{P}_1[
\{i \textrm{ visited } k \textrm{ times} \}
|\{T_L<T_0\}
]
\end{displaymath}
and, using the definition of conditional probability and 
the Markov property, 
\begin{displaymath}
\begin{split}
&
\mathbb{E}_1[\tau_i|T_L<T_0]
\\
&
\phantom{m}
=
\sum_{k=1}^\infty k 
(\mathbb{P}_1[\{T_i<T_0\}]/\mathbb{P}_1[\{T_L<T_0\}])\\
&
\phantom{mmmmm}
\times
\mathbb{P}_i[\{T_L<T_0\}
\cap\{k-1 \textrm{ returns to } i\}]
\;.
\end{split}
\end{displaymath}
The last probability appearing in the above expression is nothing but the 
quantity $Q_k$ defined for 
the five state chain with the jump 
probabilities defined as in \eqref{dimo050}.
Finally, \eqref{dimo040} follows by noting that 
\begin{displaymath}
\sum_{k=1}^\infty k Q_k
=
\frac{p_Bp_C}{[1-(r_B+q_Bp_A+p_Bq_C)]^2}
\;\;.
\end{displaymath}

Our strategy to compute the residence time is the following: 
for any $i=1,\dots,L-1$ we shall compute 
$\mathbb{E}_1[\tau_i|T_L<T_0]$ identifying the correct 
values of $p_A$, $q_A$, $p_B$, $q_B$, $p_C$, and $q_C$ to be used,
whose definition depends on the choice of the site $i$. Finally, 
the sum \eqref{dimo030} will provide us with the 
residence time. 

\subsection{Residence time in the symmetric case}
\label{s:dimoS} 
In the symmetric case, that is $\epsilon=0$ and $\lambda=1/2$,
by using the gambler's ruin 
result
we have that
\begin{equation}
\label{dimoS000}
\mathbb{P}_1[T_0<T_L]
=
\frac{L-1}{L}
\end{equation}
and
\begin{equation}
\label{dimoS010}
\mathbb{P}_1[T_L<T_0]
=
\frac{1}{L}
\;\;.
\end{equation}
This is a very classical problem in probability theory which 
can be found in any probability text book, see, for example, 
\cite[paragraphs~2 and 3, Chapter~XIV]{Fbook1968}.

\begin{figure}[t]
\vspace{-2.5 cm}
\centerline{%
\hspace{3.7 cm}
{\includegraphics[width=.9\textwidth]{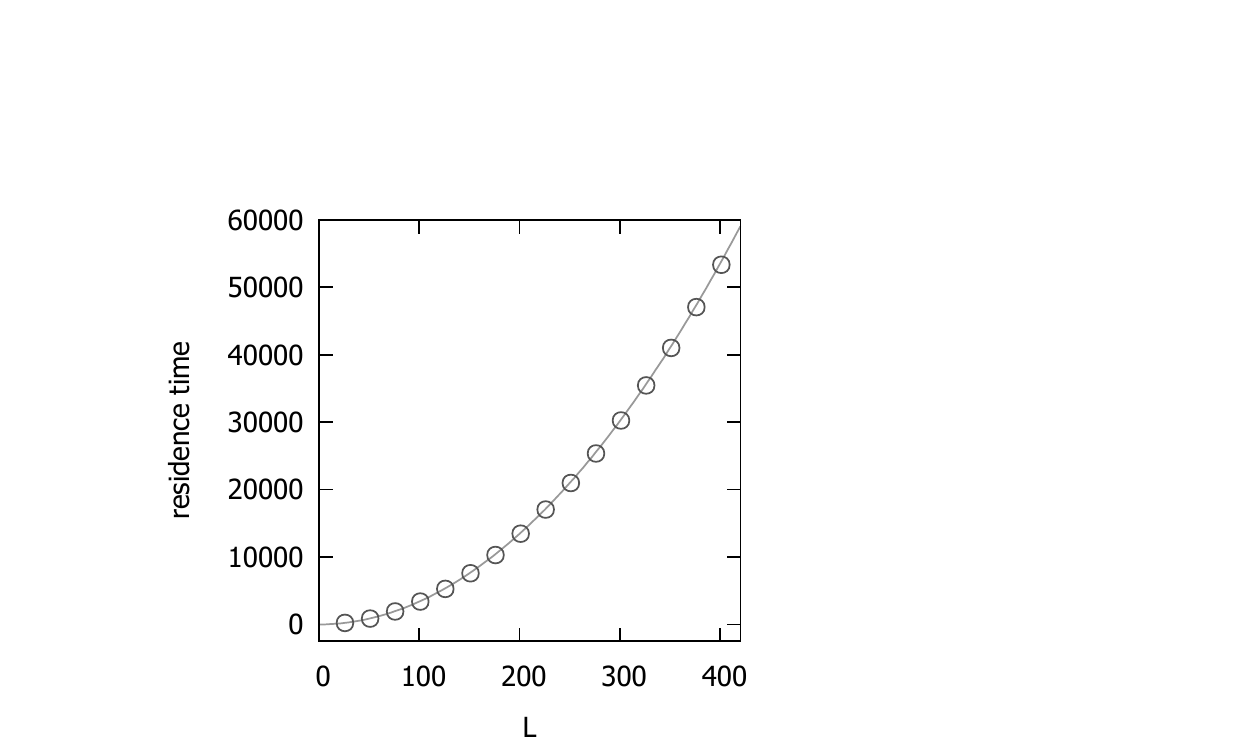}}
}
\caption{
Residence time vs.\ $L$ for the lane with no singular sites (symmetric 
case).
The solid line is the exact solution \eqref{dimoS020},
whereas circles are the average results of a Monte Carlo simulation
with $5\cdot10^8$ particles started at the site $1$.}
\label{f:sim}
\end{figure}

The computation of the residence time, which, in the gambler language, 
is the average duration of the game conditioned to the fact that the 
gambler wins, is not immediate. 
We use the formulas \eqref{dimo030}--\eqref{dimo050} proven above 
by defining suitably the five state chain jump probabilities. 
More precisely, 
$p_A=(i-1)/i$ is given by \eqref{dimoS000} with the initial point 
$1$ replaced by $i-1$ and $L$ replaced by $i$, 
$q_A=1/i$ is similarly given by \eqref{dimoS010},
$p_B=q_B=1/2$ (and hence $r_{B}=0$), 
$p_C=1/(L-i)$ is given by \eqref{dimoS010} with the initial point $1$ replaced 
by $i+1$ and $L$ replaced by $L-i$, 
and 
$q_C=(L-i-1)/(L-i)$ is given similarly by \eqref{dimoS000}.
Moreover, since from \eqref{dimoS000} it also follows that 
$\mathbb{P}_1[T_L<T_0]=1/L$ and 
$\mathbb{P}_1[T_i<T_0]=1/i$, from \eqref{dimo040} a straightforward 
computation yields 
\begin{displaymath}
\mathbb{E}_1[\tau_i|T_L<T_0]
=
\frac{2}{L}(Li-i^2)
\end{displaymath}
and, computing the sum in \eqref{dimo030}, we finally have 
\begin{equation}
\label{dimoS020}
R=
\frac{1}{3}(L-1)(L+1)
\;\;.
\end{equation}
In figure~\ref{f:sim} the numerical estimate of the residence time 
in this symmetric case is compared to the exact result \eqref{dimoS020}.
It is interesting to remark that 
the mean time 
that a symmetric walk started at $0$ needs to reach either 
$-L$ or $+L$ is $L^2$. This time can be 
computed as the average duration of the gambler's game.
Thus, conditioning the particle to exit through 
the right end point decreases by a multiplicative 
factor the mean time that the particle needs to 
reach the distance $L$ from the starting point, but it does not change 
the diffusive dependence on the length $L$ of the lane. 

\subsection{Crossing probability in the general case}
\label{s:dimoC} 
We now come back to the general 1D model introduced in Section~\ref{s:1D}.
As a first step in the residence time computation, we have to calculate
the \emph{crossing probability}
$\mathbb{P}_1[T_L<T_0]$ which appears at the denominator in 
\eqref{dimo040}.
We first note that, 
by using repeatedly the Markov property, one gets
\begin{equation}
\label{dimoC000}
\mathbb{P}_1[T_0<T_L]
=
1-p_1p_2p_3p_4p_5
\end{equation}
and, as a consequence
\begin{equation}
\label{dimoC010}
\mathbb{P}_1[T_L<T_0]
=
p_1p_2p_3p_4p_5
\end{equation}
where
\begin{displaymath}
\begin{split}
&
p_1=\mathbb{P}_1[T_0>T_{n+1}],\;
p_2=\mathbb{P}_{n+1}[T_0>T_{n+2}],\;
\\
&
p_3=\mathbb{P}_{n+2}[T_0>T_{n+h+2}],\;
p_4=\mathbb{P}_{n+h+2}[T_0>T_{n+h+3}],\;
\\
&
p_5=\mathbb{P}_{n+h+3}[T_0>T_L].
\end{split}
\end{displaymath}

The probabilities $p_1,\dots,p_5$ can be computed explicitly 
and the remaining part of this section is devoted to the 
computation of these quantities. 
For $p_1$ one has to use \eqref{dimoS010} with $L$ replaced by $n+1$ 
to deduce that 
\begin{equation}
\label{dimoC020}
p_1=\frac{1}{n+1}
\;\;.
\end{equation}

To compute $p_2$, we first note that, once the particle is in $n$, 
the probability to come back to $n+1$ before reaching $0$ is equal 
to $n/(n+1)$, as it follows by using \eqref{dimoS000} with the initial 
point $1$ replaced by $n$ and $L$ replaced by $n+1$.  Hence,
\begin{displaymath}
p_2=
\sum_{r=0}^\infty\sum_{k=0}^{r}
{{r}\choose{k}} 
\Big[(1-\varepsilon-\lambda)\frac{n}{n+1}\Big]^{r-k}
\varepsilon^k\lambda
\end{displaymath}
where $r$ counts the number of times that, starting from $n+1$, 
the walker either jumps to $n$ or it stays in $n+1$ and 
$k$ counts the number of times that the walker stays in $n+1$. 
Using the binomial theorem, we get 
\begin{equation}
\label{dimoC030}
p_2=\frac{\lambda}{1-[(1-\varepsilon-\lambda)n/(n+1)+\varepsilon]}
\;.
\end{equation}

In order to compute $p_3$, note that, using \eqref{dimoS000} and 
\eqref{dimoS010} with initial point $n+2$ and replacing $L$ with 
$h+1$, one has 
$\mathbb{P}_{n+2}[T_{n+1}<T_{n+h+2}]=h/(h+1)$ and 
$\mathbb{P}_{n+2}[T_{n+h+2}<T_{n+1}]=1/(h+1)$. Hence, 
\begin{equation}
\label{dimoC040}
p_3=
\frac{1}{h+1}
\sum_{k=0}^\infty
\Big(\frac{h}{h+1}\Big)^k p_2^k
=
\frac{1}{1+h(1-p_2)}
\;,
\end{equation}
where $k$ counts the number of times that, starting from $n+2$, 
the walker reaches $n+1$ before $n+h+2$. 

To compute $p_4$, we first need to calculate 
$\xi=\mathbb{P}_{n+h+1}[T_0>T_{n+h+2}]$. Starting from $n+h+1$ 
the probability 
to reach $n+h+2$ before $n+1$ is 
$\mathbb{P}_{n+h+1}[T_{n+h+2}<T_{n+1}]=h/(h+1)$, where we used 
\eqref{dimoS000} with initial point $n+h+1$ and $L$ replaced by $h+1$.  
Hence, 
$\mathbb{P}_{n+h+1}[T_{n+1}<T_{n+h+2}]=1/(h+1)$.
Thus, 
\begin{displaymath}
\xi
=
\frac{h}{h+1}
+
\frac{1}{h+1}
p_2
\frac{1}{h+1}
 \sum_{k=0}^\infty\Big(p_2\frac{h}{h+1}\Big)^k
\end{displaymath}
where $k$ counts the number of times that the walker returns to 
$n+1$ after having visited it for the first time. We have also 
used that 
$\mathbb{P}_{n+2}[T_{n+1}<T_{n+h+2}]=h/(h+1)$.
With some algebra we find the expression
\begin{equation}
\label{dimoC050}
\xi=\frac{p_2+h(1-p_2)}{1+h(1-p_2)}
\;\;.
\end{equation}
Now, we have all the ingredients to compute $p_4$. 
Indeed, 
\begin{displaymath}
p_4=
(1-\varepsilon-\lambda)
\Big[
\sum_{r=0}^\infty 
\sum_{k=0}^{r}
{{r}\choose{k}}
\varepsilon^k (\lambda\xi)^{r-k}
\Big]
\end{displaymath}
where $r-k$ counts the number of times that the walker starting from 
$n+h+2$ jumps to $n+h+1$ and
where $k$ counts the number of times that the walker stays at 
$n+h+2$.
A simple calculation provides the result 
\begin{equation}
\label{dimo060}
p_4=\frac{1-\varepsilon-\lambda}{1-(\lambda\xi+\varepsilon)}
\;\;.
\end{equation}

Finally, to compute $p_5$ we remark that 
$\mathbb{P}_{n+h+3}[T_L<T_{n+h+2}]=1/(w+1)$ and 
$\mathbb{P}_{n+h+3}[T_{n+h+2}<T_L]=w/(w+1)$, as it can be deduced by 
\eqref{dimoS010} and \eqref{dimoS000} by using as initial point the point 
$n+h+3$ and replacing $L$ by $w+1$. 
Then, 
\begin{equation}
\label{dimoC060}
p_5
=
\frac{1}{w+1}
\sum_{k=0}^\infty
\Big(\frac{w}{w+1}p_4\Big)^k
=\frac{1}{1+w(1-p_4)}
\;\;.
\end{equation}

Finally, plugging the equations \eqref{dimoC020}--\eqref{dimoC060}
in \eqref{dimoC010}, we find the expression 
\begin{equation}
\label{dimoC070}
\mathbb{P}_1[T_L<T_0]
=
\frac{\lambda}{(1+h)(1-\varepsilon-2\lambda)+\lambda L}
\end{equation}
for the probability that the particle started at the site $1$ 
reaches $L$ before visiting $0$. It is interesting to remark that 
in the case $\varepsilon=0$ and $\lambda=1/2$ the 
expression \eqref{dimoS010} valid in the symmetric case is recovered.

\subsection{Residence time in presence of defects}
\label{s:dimoR} 
The last step, necessary to complete our algorithm to 
compute the residence time, is that of listing the expression that 
must be used for the probabilities \eqref{dimo050} for the 
different choices of $i$ on the lattice. 
In this last section, in order to get simpler formulas, we focus on the 
case that has been studied numerically, that is to say, 
we choose the parameterization \eqref{one020}.
First of all we note that the expression \eqref{dimoC070} of the 
probability that the particle started at the site $1$ 
reaches $L$ before visiting $0$ simplifies to 
\begin{equation}
\label{dimoR000}
\mathbb{P}_1[T_L<T_0]
=
\frac{1-p}{p(1+h)+(1-p)L}
\;\;.
\end{equation}

The site $i$ in the lattice can be chosen in nine 
possible different ways: in the bulk of the three regions 
on the left, between and on the right of the defect sites, 
as one of the four sites neighboring the defects and as one of the 
two defect site.
We list only five cases, the remaining four can be deduced
exchanging the role of the parameters $n$ and $w$. 
Note that we shall only list either $p_A$ or $q_A$ 
and $p_C$ or $q_C$; the missing parameter can be deduced 
by the equations 
$p_A+q_A=1$ and $p_C+q_C=1$.

\smallskip
\par\noindent{\textit{Case $1\le i\le n-1$.\/}
First note that $\mathbb{P}_1[T_i<T_0]=1/i$ is given by  
\eqref{dimoS010} with initial site $1$ and $L$ replaced by $i$. 
Moreover, 
$p_A=(i-1)/i$ follows from \eqref{dimoS000} with initial site $i-1$ and 
$L$ replaced by $i$. 
We trivially have that $p_B=q_B=1/2$.
Finally, $p_C=(1-p)/[p(1+h)+(1-p)(L-i)]$ follows from 
\eqref{dimoR000} with initial site $i+1$ and $L$ replaced by 
$L-i$.

\smallskip
\par\noindent{\textit{Case $i=n$.\/}
First note that $\mathbb{P}_1[T_i<T_0]=1/n$ is given by  
\eqref{dimoS010} with initial site $1$ and $L$ replaced by $n$. 
Moreover, 
$p_A=(n-1)/n$ follows from \eqref{dimoS000} with initial site $n-1$ and 
$L$ replaced by $n$. 
We trivially have that $p_B=q_B=1/2$.
Finally, we note that $q_C$ has the same structure as 
$p_4$, thus, by exchanging the role of $n$ and $w$, from 
\eqref{dimoC030}, \eqref{dimoC050}, and \eqref{dimoC060} 
we have that 
$q_C=1/[2-p-(1-p)\zeta]$
where
\begin{equation}
\label{dimoR010}
\zeta=\frac{\pi+h(1-\pi)}{1+h(1-\pi)}
\;\textrm{ and }\;
\pi=\frac{1-p}{2-p-\frac{w}{w+1}}
\;.
\end{equation}

\smallskip
\par\noindent{\textit{Case $i=n+1$.\/}
First note that $\mathbb{P}_1[T_i<T_0]=1/(n+1)$ is given by  
\eqref{dimoS010} with initial site $1$ and $L$ replaced by $n+1$. 
Moreover, 
$p_A=n/(n+1)$ follows from \eqref{dimoS000} with initial site $n$ and 
$L$ replaced by $n$. 
We trivially have that $p_B=(1-p)/2$ and $q_B=1/2$.
Finally, we note that $q_C$ has the same structure as 
$\xi$, thus, by exchanging the role of $n$ and $w$, from 
\eqref{dimoC050}
we have that 
$q_C=\zeta$, see \eqref{dimoR010}.

\smallskip
\par\noindent{\textit{Case $i=n+2$.\/}
First note that $\mathbb{P}_1[T_i<T_0]=p_1p_2$, hence, using 
\eqref{dimoC020} and \eqref{dimoC030}, an easy 
computation yields
$\mathbb{P}_1[T_i<T_0]=(1-p)/[(n+1)(2-p-n/(n+1))]= (1-p)/(2+n-p(n+1))$ since, 
with the parameterization that we are adopting in this section
\begin{displaymath}
p_2=\frac{1-p}{2-p-n/(n+1)}
\;\;.
\end{displaymath}
Moreover, $p_A=p_2$ by definition 
and $p_B=q_B=1/2$.
Finally, we note that $q_C$ has the same structure as 
$\xi$ with $h$ replaced by $h-1$. Thus, by exchanging the role of 
$n$ and $w$, from 
\eqref{dimoC050}
we have that 
$q_C=[\pi+(h-1)(1-\pi)]/[1+(h-1)(1-\pi)]$, 
with $\pi$ defined in \eqref{dimoR010}.

\smallskip
\par\noindent{\textit{Case $n+3\leq i \leq n+h$.\/}
First note that $\mathbb{P}_1[T_i<T_0]=p_1p_2 \bar{p}_3$, where $\bar{p}_3$ has the structure of $p_3$ with $h$ replaced by $i-(n+2)$.  Hence 
\eqref{dimoC040} 
gives us
$\mathbb{P}_1[T_i<T_0]= (p_1p_2)/(1 +( i-n-2)(1-p_2)) $
with $p_1$ and $p_2$ as in the previous case.
Moreover, $p_A$ has the same structure as 
$\xi$ with $h$ replaced by $i-n-2$ so $
p_A=({p_2+(i-n-2)(1-p_2)})/({1+(i-n-2)(1-p_2)})$
and $p_B=q_B=1/2$.
Finally, we note that $q_C$ has the same structure as 
$\xi$ with $h$ replaced by $n+h+1-i$. Thus, by exchanging the role of 
$n$ and $w$, from 
\eqref{dimoC050}
we have that 
$q_C=[\pi+(n+h+1-i)(1-\pi)]/[1+(n+h+1-i)(1-\pi)]$, 
with $\pi$ defined in \eqref{dimoR010}.

\section{Conclusions}
\label{s:conclusioni} 
\par\noindent
We have studied in detail the effect of an obstacle in a 
2D strip on the flux of particles performing a simple symmetric 
random walk. We have found that, due to purely geometrical effects,
the typical time that a particle entered in strip through the 
left boundary and leaving the system through the right boundary
has a complex dependence on the geometrical parameters of the obstacle. 
In particular, we stress that we found non--monotonic behaviors as 
a function of a sufficiently large obstacle.
These phenomena have been interpreted in terms of the total time that 
the particles spend in each of the three regions of the strip in which 
the obstacle naturally partitions the lattice: the one on its left, the 
one on its right, and the channels between the obstacle and the horizontal 
boundary. 
Finally, we have studied numerically and analytically a 1D model 
mimicking the 2D random walk and we have found similar results. 
In this case we have been able to develop a complete analytical 
computation and to compare our numerical results to the exact 
solution. 


\begin{acknowledgments}
ENMC thanks R.\ van der Hofstad for very useful discussions. 
\end{acknowledgments}



\end{document}